\documentclass[aps,prl,twocolumn,superscriptaddress,showpacs,floatfix,a4paper]{revtex4}
\usepackage{graphicx}
\usepackage[intlimits]{amsmath}

\allowdisplaybreaks[1]

\begin{document}

\title{Population switching and charge sensing in quantum dots:
A case for a quantum phase transition}

\author{Moshe Goldstein}
\author{Richard Berkovits}
\affiliation{The Minerva Center, Department of Physics, Bar-Ilan
University, Ramat-Gan 52900, Israel}

\author{Yuval Gefen}
\affiliation{Department of Condensed Matter Physics,
The Weizmann Institute of Science, Rehovot 76100, Israel}

\begin{abstract}
A broad and a narrow level of a quantum dot connected to two
external leads may swap their respective occupancies as a function
of an external gate voltage. By mapping this problem onto 
a multi-flavored Coulomb-gas we show that such population
switching is not abrupt.
However, trying to measure it by adding a third
electrostatically coupled lead may render this switching an abrupt
first order quantum phase transition. This is related to the
interplay of the Mahan mechanism versus the Anderson orthogonality
catastrophe, in similitude to the Fermi edge singularity.
A concrete setup for experimental observation of this effect
is also suggested.
\end{abstract}


\maketitle

\begin{figure}
\includegraphics[width=!,height=4cm]{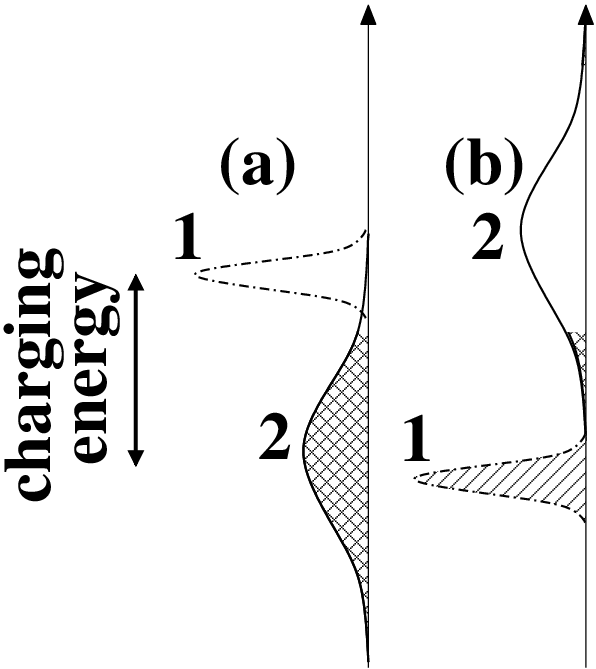}
\caption{\label{fig:popswitch}
Respective occupation of levels 1 and 2 (a) before and (b) after
population switching has taken place.
}
\end{figure}

The phenomenon of \emph{population switching} (PS)
\cite{population_switching,berkovits05,ps_theory}
occurs in a discrete level quantum dot (QD) --- e.g., a QD with one
broad and one narrow level. 
Upon a continuous variation of a plunger gate voltage the occupation
numbers of the levels are inverted, cf.\ Fig.~\ref{fig:popswitch}.
This phenomenon is relevant to a wide range
of experimentally observed effects, including the widely used technique of
\emph{charge sensing} \cite{charge_sensing}
and, possibly, the occurrence of a large shot noise Fano factor
through a QD \cite{zarchin07}, as well as correlated lapses
\cite{phase_lapse} of the transmission phase through a QD
\cite{pl_theory,slava}.
One particularly intriguing question in this context is whether or not
(at zero temperature) PS is abrupt, and hence constitutes a first order
quantum phase transition (QPT).

\begin{figure}
\includegraphics[width=5.5cm,height=!]{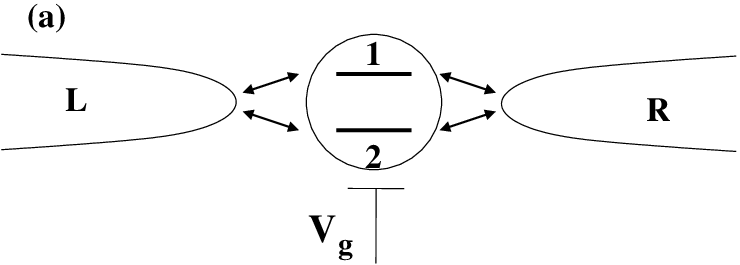}
\includegraphics[width=6.5cm,height=!]{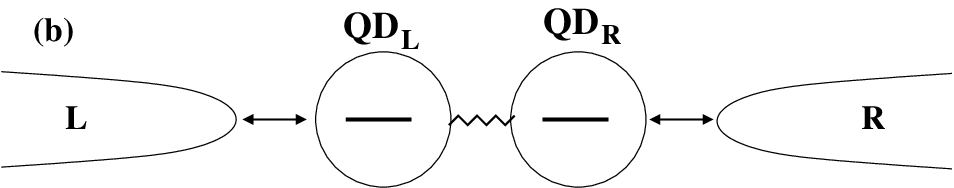}
\vskip 0.1cm
\includegraphics[width=6.5cm,height=!]{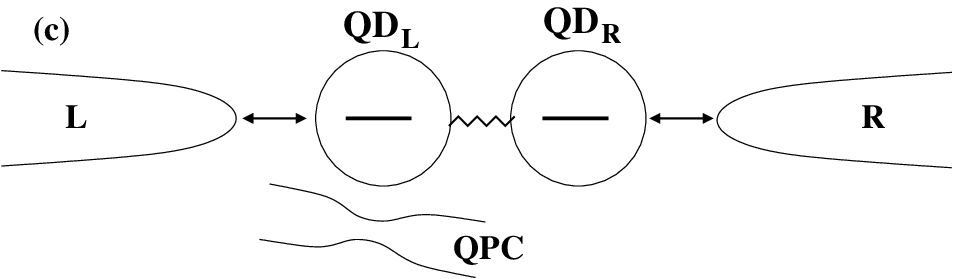}
\vskip -0.3cm
\includegraphics[width=6.5cm,height=!]{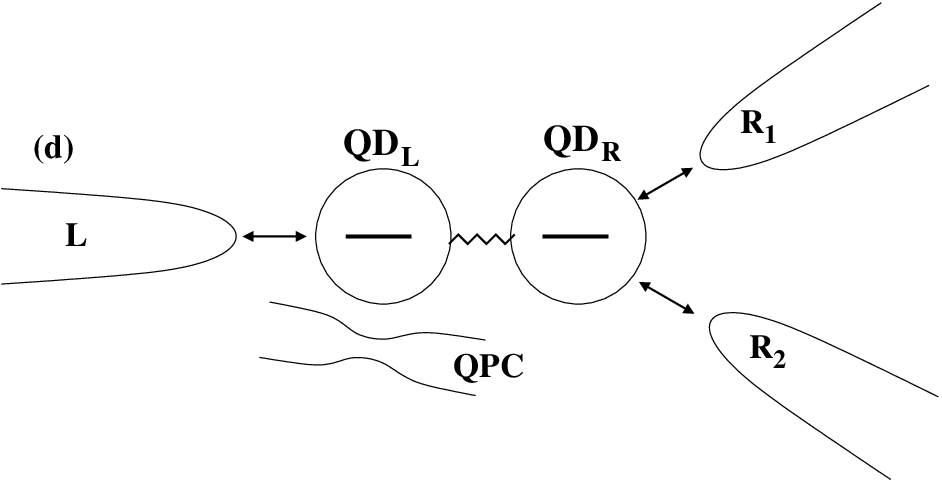}
\caption{\label{fig:geometry}
The systems discussed:
(a) the original model of a two-level quantum dot;
(b) an equivalent model of two electrostatically-coupled
single-level QDs;
(c) a third terminal (QPC) added;
(d) right level tunnel-coupled to two leads.}
\end{figure}

In the following we address this question in the context of a two
level QD coupled to leads of spinsless noninteracting electrons.
This is mapped onto a system of two single level QDs,
each coupled to a single lead (cf.\ Fig.~\ref{fig:geometry}). We
formulate the problem in terms of a multi-flavored Coulomb-gas (CG),
perform a renormalization group (RG) analysis of this 15
parameter problem, and show that its low temperature behavior is
akin to an antiferromagnetic Kondo problem, hence no QPT occurs.
This is dramatically changed when a third lead
(e.g., a quantum point contact, QPC, serving as a charge sensor)
is electrostatically coupled to one of the QDs. The
model may then scale to the ferromagnetic Kondo problem, and by
tuning the strength of the third lead coupling one induces a QPT.

The problem at hand can be viewed within an even broader context.
The features of the Fermi edge singularity are the
result of the competition between the Anderson orthogonality
catastrophe and the Mahan exciton physics \cite{mahan}. The fact
that the latter wins gives rise to the divergence at the X-ray
absorption edge. Such an interplay is present here too. Turning on
the electrostatic coupling to the third lead increases the weight of the
orthogonality catastrophe. The latter eventually wins, suppressing
transitions between charge configurations before and after PS takes
place. This implies a QPT between these two configurations.
Our setup then serves as a handy laboratory which allows us to
control and tune the relative strengths of two fundamental effects
in many-body physics.

The original system of spinless electrons, made of a two (unequal)
orbital level QD connected to two leads
[cf.\ Fig.~\ref{fig:geometry}(a)],
is described by the Hamiltonian $H =
\tilde{H}_\text{lead} + \tilde{H}_\text{dot} + \tilde{H}_\text{dot-lead}$.
We assume the leads to be non-interacting,
and the dot-lead tunneling matrix elements
$\tilde{V}_{i \ell}$
($i=1,2$ and $\ell = L,R$ for left, right)
to be real and possess left-right symmetry,
$| \tilde{V}_{i L} | = | \tilde{V}_{i R} |$
(effects of asymmetry are discussed later).
We will consider the more intricate case
$\text{sgn} ( \tilde{V}_{1L} \tilde{V}_{1R} \tilde{V}_{2L} \tilde{V}_{2R} )
= -1$
\cite{pl_theory}.
We now map the original model onto a modified one
consisting of two single-level QDs, cf.\ Fig.~\ref{fig:geometry}(b).
The Fermi operators $\psi_L$ and $\psi_R$ of the new leads are made of symmetric
and antisymmetric combinations of the original
$\tilde{\psi}_L$ and $\tilde{\psi}_R$, respectively.
The Hamiltonian is $H = \sum_{\ell} H_\ell + H_U$, 
with $H_\ell = H_{\ell,\text{lead}} + H_{\ell,\text{dot}} + H_{\ell,\text{dot-lead}}$
where $H_{\ell,\text{lead}}$ describes the Fermi liquid of the respective lead,
$H_{\ell,\text{dot}} = \varepsilon_\ell d_\ell^\dagger d_\ell$
($d^\dagger_\ell$ is the creation operator at dot $\ell$),
$H_{\ell,\text{dot-lead}} = V_\ell d_\ell^\dagger \psi_\ell(0) + \text{H.c.}$
with
$V_L = \sqrt{2} |\tilde{V}_{1L}|$ and $V_R = \sqrt{2} |\tilde{V}_{2R}|$,
and, finally, $H_U = U d_L^{\dagger} d_L d_R^\dagger d_R$.
The dot-lead coupling gives rise to level widths
$\Gamma_\ell = \pi |V_\ell|^2 \rho_\ell$,
where $\rho_\ell$ are the local densities of states.

\begin{figure}
\includegraphics[width=!,height=3.5cm]{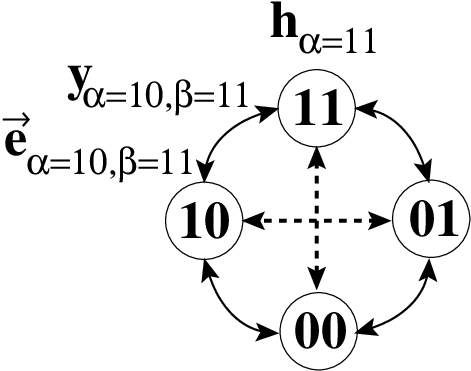}
\caption{\label{fig:cg_params}
Illustration of the parameters characterizing the CG 
[Eqs.~(\ref{eqn:cg_cardy1})--(\ref{eqn:cg_cardy2})]:
$h_{\alpha=11}/\xi$ is the energy associated with the state 11;
its bare value is $(\varepsilon_L+\varepsilon_R+U)$. The transition
depicted involves the fugacity $y_{\alpha=10,\beta=11}$ and the
charge $\vec{e}_{\alpha=10,\beta=11}$, whose bare values are
$\sqrt{\Gamma_R \xi / \pi}$ and $(0,1)$, respectively (see
Table~\ref{tbl:cg_params}). Dashed lines indicate couplings
generated through RG iterations, e.g., $y_{10,01}$
(corresponding to $\rho J_{xy}/2$).
}
\end{figure}

Using standard techniques, we can rewrite the partition function as
that of a classical multi-flavor one-dimensional CG
\cite{yuval_anderson,cardy81,kamenev_gefen,long_version}.
The imaginary time history of the system (
a circle of circumference $1/T$, the inverse temperature)
is divided into intervals in which
the system is in one of four states spanning the filling
configurations of the two dots: $\alpha=00$, $10$, $01$, and $11$
(cf.\ Fig.~\ref{fig:cg_params}).
The state $\alpha$ has dimensionless energy $h_\alpha$.
These intervals are separated by transition events, the CG particles.
A transition from configuration $\alpha$ to $\beta$ ($\alpha \ne \beta$)
is associated with a fugacity $y_{\alpha \beta} = y_{\beta \alpha}$,
and a two-component vector charge
$\vec{e}_{\alpha \beta} = -\vec{e}_{\beta \alpha}$
(the two components correspond to the
charge removed from the $L$ and $R$ lead, respectively,
cf.\ Table~\ref{tbl:cg_params}).
The partition function reads: 
\begin{equation}
\label{eqn:cg_cardy1}
 Z = \sum_{N=0}^{\infty} \sum_{\alpha_i, \beta_i}
 y_{\alpha_1 \beta_1}
 \dots y_{\alpha_N \beta_N} 
 \negthickspace \negthickspace
 \int_0^{1/T} \negthickspace \frac{\text{d} \tau_{2N}}{\xi}
 \dots \negthickspace \negthickspace 
 \int_0^{\tau_2 - \xi} \negthickspace \frac{\text{d} \tau_1}{\xi} 
 \text{e}^{- S ( \{ \tau_i, \alpha_i \})} 
\end{equation}
where $\beta_i = \alpha_{i+1}$, $N+1 \equiv 1$,
and $\xi$ is a short-time cutoff.
The classical CG action is:
\begin{equation}
\label{eqn:cg_cardy2}
 S( \{ \tau_i, \alpha_i \} ) =
 \sum_{i<j=1}^N \vec{e}_{\alpha_i \beta_i} \cdot \vec{e}_{\alpha_j \beta_j}
 V_C (\tau_j-\tau_i)
 + \sum_{i=1}^N h_{\beta_i} \frac{\tau_{i+1} - \tau_{i}}{\xi}.
\end{equation}
with $V_C(\tau-\tau^\prime) =
\ln \left\{ \pi T \xi / \sin [ \pi T (\tau-\tau^\prime) ] \right\}$.
Bare values of the CG parameters are listed in
Table~\ref{tbl:cg_params}.

We can now write down a set of 15 RG equations
for the CG parameters
(valid to second order in the fugacities but otherwise exact; here
$\kappa_{\alpha \beta} \equiv \left| \vec{e}_{\alpha \beta}
\right|^2$ and $\kappa^{\alpha}_{\beta \gamma} \equiv \kappa_{\alpha
\beta} + \kappa_{\alpha \gamma} - \kappa_{\beta \gamma}$) 
\cite{yuval_anderson,cardy81}: 
\begin{align}
 \label{eqn:rg1}
 \frac{d y_{\alpha \beta}}{d \ln \xi} = &
 \frac{2-\kappa_{\alpha \beta}}{2}
 y_{\alpha \beta}
 + \sum_{\gamma} y_{\alpha \gamma}y_{\gamma \beta}
    e^{ (h_{\alpha} + h_{\beta})/2 - h_{\gamma} }, \\
 \label{eqn:rg2}
 \frac{d \kappa_{\alpha \beta}}{d \ln \xi} = &
 - \sum_{\gamma} y_{\alpha \gamma}^2 e^{ h_{\alpha} - h_{\gamma} }
    \kappa^{\alpha}_{\beta \gamma}
 - \sum_{\gamma} y_{\beta \gamma}^2 e^{ h_{\beta} - h_{\gamma} }
    \kappa^{\beta}_{\alpha \gamma}, \\
 \label{eqn:rg3}
 \frac{d h_{\alpha}}{d \ln \xi} = & h_{\alpha}
 - \sum_{\gamma} y_{\alpha \gamma}^2 e^{ h_{\alpha} - h_{\gamma} }
 + \frac{1}{4} \sum_{\beta, \gamma} y_{\beta \gamma}^2 e^{ h_{\beta} - h_{\gamma} }.
\end{align}

\begin{table}
\caption{ \label{tbl:cg_params}
Parameters appearing in the CG expansion 
[Eqs.~(\ref{eqn:cg_cardy1})--(\ref{eqn:cg_cardy2})],
corresponding to the system depicted in Fig.~\ref{fig:geometry}(b).
}
\begin{ruledtabular}
\begin{tabular}{ccc}
 Fugacities & Charges & Energies \\ \hline
 $y_{00,10} = \sqrt{\Gamma_L \xi / \pi}$ & 
 $\vec{e}_{00,10} = \left( 1, 0 \right)$ &
 $h_{00} = 0$ \\
 $y_{00,01} = \sqrt{\Gamma_R \xi / \pi}$ & 
 $\vec{e}_{00,01} = \left( 0, 1 \right)$ &
 $h_{10} = \varepsilon_L \xi$ \\
 $y_{10,11} = \sqrt{\Gamma_R \xi / \pi}$ & 
 $\vec{e}_{10,11} = \left( 0, 1 \right)$ &
 $h_{01} = \varepsilon_R \xi$ \\
 $y_{01,11} = \sqrt{\Gamma_L \xi / \pi}$ & 
 $\vec{e}_{01,11} = \left( 1, 0 \right)$ &
 $h_{10} = ( \varepsilon_L + \varepsilon_R + U) \xi$ \\
 $y_{10,01} = 0$ &
 $\vec{e}_{10,01} = \left( -1, 1 \right)$ \\
 $y_{00,11} = 0$ &
 $\vec{e}_{00,11} = \left( 1, 1 \right)$ \\
\end{tabular}
\end{ruledtabular}
\end{table}

We now address the parameter regime in the vicinity of population switching.
This requires a small level separation,
$|\varepsilon_L - \varepsilon_R| < |\Gamma_L - \Gamma_R|$.
Defining $\varepsilon_0 = (\varepsilon_L + \varepsilon_R) / 2$, we have,
in the Coulomb-blockade valley,
$|\varepsilon_0|, \varepsilon_0+U \gg \Gamma_L, \Gamma_R$.
The RG flow is then divided into three stages:
(i) $\xi^{-1} \gg \max( |\varepsilon_0|, \varepsilon_0+U ) $,
all four filling configurations take equal part in the RG flow;
(ii) $\min( |\varepsilon_0|, \varepsilon_0+U ) \ll \xi^{-1} \ll
\max( |\varepsilon_0|, \varepsilon_0+U ) $,
the higher energy configuration between $00$ and $11$ drops out;
(iii) $\xi^{-1} \ll \min( |\varepsilon_0|, \varepsilon_0+U )$, only
configurations $10$ and $01$ survive.
In this last stage we are left with a CG 
of only a single type of transitions --- equivalent to
the single channel anisotropic Kondo model \cite{yuval_anderson}.
The main effect of the first two stages of the flow is to
establish the fugacity of the 10$\rightleftharpoons$01 transition,
(via virtual processes through the doubly occupied and unoccupied states,
11 and 00),
as well as to renormalize the corresponding charge
and the energy difference between these states.

The resulting Kondo model has the following
couplings, to leading order in $\Gamma_\ell$
(all the parameters refer to their bare values)
\cite{long_version}:
\begin{widetext}
\begin{align}
 \label{eqn:jz}
 \rho J_z = & 1 - \frac{\kappa_{10,01}}{2} 
    + \left[ \frac{\Gamma_L}{2\pi} \left(
    \frac{Q_{2\kappa_{00,10}}(|\varepsilon_L| \xi)} {|\varepsilon_L|}
        \kappa^{10}_{01,00}
    + \frac{Q_{2\kappa_{01,11}} ( [ \varepsilon_L + U ] \xi )}
        {\varepsilon_L + U} \kappa^{01}_{10,11}
    \right)
    + \left\{ L \leftrightarrow R , 10 \leftrightarrow 01 \right\} \right]
    , \\
 \label{eqn:jxy}
 \rho J_{xy} = & \frac{2\sqrt{ \Gamma_L \Gamma_R} } {\pi}
    \left[
    \frac{Q_{\kappa_{00,10}+\kappa_{00,01}}
        (|\varepsilon_0| \xi)} {|\varepsilon_0|}
    + \frac{Q_{\kappa_{10,11}+\kappa_{01,11}}
        ( [ \varepsilon_0 + U ] \xi)}{\varepsilon_0 + U}
    \right], \\
 \label{eqn:hz}
 H_z = & \varepsilon_L - \varepsilon_R 
    - \frac{\Gamma_L}{\pi} \left[
    P_{2\kappa_{00,10}}(|\varepsilon_L| \xi)
    - P_{2\kappa_{01,11}} ( [ \varepsilon_L + U ] \xi )
    \right] 
    + \frac{\Gamma_R}{\pi} \left[
    P_{2\kappa_{00,01}}(|\varepsilon_R| \xi)
    - P_{2\kappa_{10,11}} ( [ \varepsilon_R + U ] \xi )
    \right] ,
\end{align}
\end{widetext}
where
$P_a(x) = \Gamma(1-a/2)/x^{1-a/2}$, $Q_a(x) = \mbox{$(1-a/2)P_a(x)$}$.
For the system discussed thus far (cf.\ Table~\ref{tbl:cg_params})
the bare values are
$\kappa_{\alpha \beta}=1$ for all $\alpha$,$\beta$
except $\kappa_{10,01} = \kappa_{00,11} = 2$. We then find:
\begin{align}
 \label{eqn:jz0}
 \rho J_z &= \frac{\Gamma_L}{\pi} \left(
    \frac{1}{\varepsilon_L + U}
    + \frac{1} {|\varepsilon_L|}
    \right) 
    + \frac{\Gamma_R}{\pi} \left(
    \frac{1}{\varepsilon_R + U}
    + \frac{1}{|\varepsilon_R|}
    \right) , \\
 \label{eqn:jxy0}
 \rho J_{xy} &= \frac{2\sqrt{ \Gamma_L \Gamma_R} } {\pi}
    \left(
    \frac{1}{\varepsilon_0 + U}
    + \frac{1}{|\varepsilon_0|}
    \right), \\
 \label{eqn:hz0}
 H_z &= \varepsilon_L - \varepsilon_R 
    - \frac{\Gamma_L}{\pi} \ln \frac{\varepsilon_L + U}{|\varepsilon_L|}
    + \frac{\Gamma_R}{\pi} \ln \frac{\varepsilon_R + U}{|\varepsilon_R|},
\end{align}
in agreement with the poor man's scaling of
Refs.~\onlinecite{slava}.
$H_z$ will change sign as the gate voltage is swept across the Coulomb blockade
valley, provided that
$|\varepsilon_L-\varepsilon_R| < |\Gamma_L-\Gamma_R|$, hence
the spin projection $\langle S_z \rangle$ will also change sign,
implying a PS.
Since $J_{xy}$ and $J_z$ are antiferromagnetic, they flow to strong coupling,
so the PS will be continuous over the scale of the
Kondo temperature,
$T_K = \frac{ \sqrt {U(\Gamma_L+\Gamma_R)} }{\pi}
\exp \left[ \frac{\pi \varepsilon_0(U+\varepsilon_0)}
{2U(\Gamma_L-\Gamma_R)} \ln \left( \frac{\Gamma_L}{\Gamma_R} \right) \right]$.

The problem becomes much more intriguing when an electrostatically
coupled third lead (e.g., a QPC charge sensor) is
introduced, cf.\ Fig.~\ref{fig:geometry}(c).
The sensor adds to the Hamiltonian a term $H_{S} =
H^{S}_\text{lead} + 
U_{S} \mbox{$:\psi_{S}^\dagger(0)\psi_{S}(0):$}%
(d_L^\dagger d_L - \frac{1}{2})$,
consisting of the Hamiltonian of a free lead
plus an interaction term. One may re-employ the CG 
formalism, but now $\vec{e}_{\alpha \beta}$ consists of three
components \cite{cardy81,long_version}. 
Denoting the population of dot $L$ in state $\alpha$ by
$n_{L\alpha}$, the third component of $\vec{e}_{\alpha \beta}$ is given by
$(n_{L \beta}-n_{L \alpha}) \delta_{S} / \pi $, with
$\delta_{S} = 2 \tan^{-1} (\pi \rho_{S} U_{S}/2)$
being the change in phase shift of the electronic wave-functions of the QPC
caused by a change in the population of dot $L$,
and $\rho_{S}$ the corresponding local density of states.
The resulting RG equations [Eqs.~(\ref{eqn:rg1})--(\ref{eqn:rg3})]
and their general solution [Eqs.~(\ref{eqn:jz})--(\ref{eqn:hz})]
are as before.
Now, however, the bare valued are
$\kappa_{00,01}=\kappa_{10,11}=1$,
$\kappa_{00,10}=\kappa_{01,11}=1+(\delta_{S}/\pi)^2$,
and
$\kappa_{00,11}=\kappa_{01,10}=2+(\delta_{S}/\pi)^2$.
At low energies we are still left with an effective Kondo model.
The main effect of the QPC would be to reduce $\rho J_z$ by
$(\delta_{S}/\pi)^2/2$, through the first term on the r.h.s.\
of Eq.~(\ref{eqn:jz}).
It may then
drive the system to the weak coupling (ferromagnetic Kondo)
regime, 
and render the PS an abrupt first order QPT.
For this to happen the QPC charge sensitivity does not need to be too high;
we require
$\rho_{S} U_{S} \sim
\left\{ \sum_\ell  \Gamma_\ell \left[
(\varepsilon_\ell + U)^{-1} + |\varepsilon_\ell|^{-1} \right] \right\}^{1/2} \ll 1$.
The transition at $J_z=-J_{xy}$ between the continuous and discontinuous PS
regimes is of the Kosterlitz-Thouless type.

Our analysis here can be put in a more general context. Around the
point where PS takes place we need to consider only
pseudo-spin up (10) and down (01) states.
Let us first ignore the QPC.
Processes in which an electron (or a hole) hops in and
out of a level (pseudo-spin diagonal, $J_z$-type processes)
give rise to an effective repulsive interaction
between the charge of each level and the charge at the
end of the nearby lead, of the form
$\sum_\ell U_\ell
\mbox{$:\psi_\ell^\dagger(0)\psi_\ell(0):$} (d^\dagger_\ell d_\ell - \frac{1}{2})$,
$U_\ell = |V_\ell|^2
\left[ (\varepsilon_\ell + U)^{-1} + |\varepsilon_\ell|^{-1} \right]$.
These correspond [by Eq.~(\ref{eqn:jz0})] to the usual $J_z S_z s_z(0)$
coupling generated in the ordinary Anderson model.
A process of the type 10$\rightleftharpoons$01 (pseudo-spin flip, $J_{xy}$ process)
contributes to the hybridization of these two configurations, hence (if relevant)
to a smearing of the PS. The aforementioned effective repulsion
introduces two competing elements into this dynamics. On the one
hand, 10$\rightleftharpoons$01 involves a change in the leads'
state, hence is suppressed by the Anderson orthogonality
catastrophe. On the other hand, an electron settling in one of the
levels has prepared itself a hole in the lead into which it can hop
(a Mahan exciton). This facilitates tunneling out and in
(by reducing Pauli-blockade), thus enhancing hybridization of
10$\rightleftharpoons$01. The overall scaling dimension is given by
$d_{xy}=1-(\delta_L + \delta_R)/ \pi + (\delta_L^2 + \delta_R^2) / 2
\pi^2$, where
$\delta_\ell = 2 \tan^{-1} ( \pi \rho_\ell U_\ell / 2 )$
is the phase-shift change in lead $\ell$ as a result of the
10$\rightleftharpoons$01 transition.
In this expression for $d_{xy}$ the linear (quadratic) term
in $\delta_\ell$ denotes the contribution of the
Mahan (Anderson) physics \cite{weichselbaum}. 
Since $\delta_\ell<\pi$, $d_{xy}<1$ (relevant), so PS is a
continuous crossover. However, when a third lead is added, the
scaling dimension is increased by $(\delta_{S}/\pi)^2/2$,
the extra orthogonality associated with the QPC
[in addition to a less important renormalization of the
the other terms, cf.\ Eqs.~(\ref{eqn:jz})-(\ref{eqn:hz})].
This may turn the Anderson effect dominant,
and the population switching abrupt. 

\begin{figure}
\includegraphics[width=!,height=7cm]{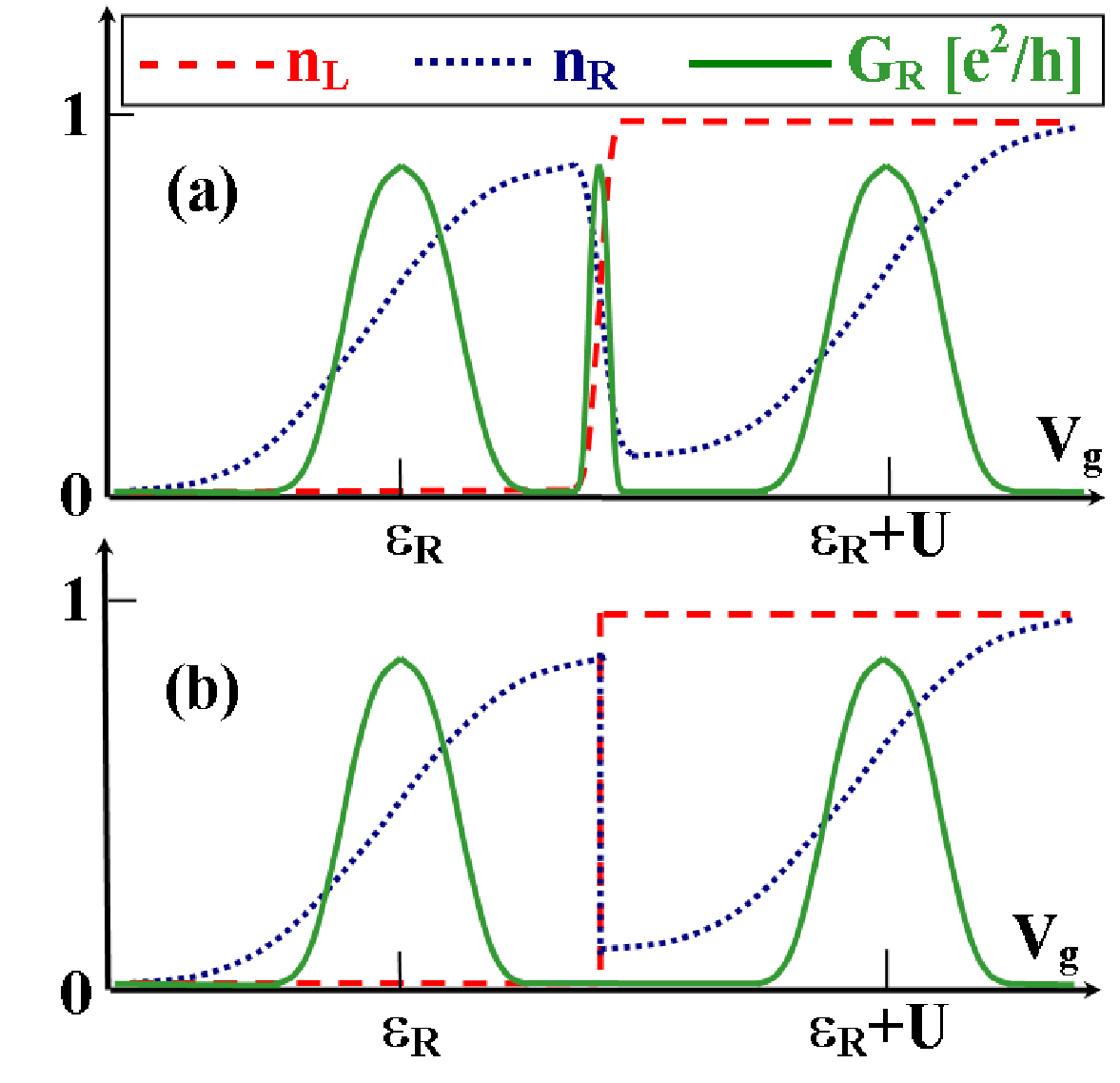}
\caption{\label{fig:friedel} (Color online) Schematic behavior of
the level populations ($n_L$ and $n_R$)
and the conductance through the right level
(taking it as the wider one) for (a) continuous or (b) abrupt
switching. See the text for further details.}
\end{figure}

Our results might raise the following concern: the QPC
can be thought of as a measuring device
(detector of the population of one QD).
At the same time, it is the introduction of this QPC which drives the
population switching from a mere crossover to a genuine QPT.
How then could one detect the difference between having and not having a
coupled QPC, i.e., between having and not having a QPT?
A solution can be achieved by attaching an additional
lead to, say, the right QD [cf.\ Fig.~\ref{fig:geometry}(d)].
Our analysis remains unchanged, provided $t_R$ is replaced by
$\sqrt{|t_{R1}|^2+|t_{R2}|^2}$, where $t_{R1(2)}$ is the tunneling
amplitude to lead $R_{1(2)}$. Now, however, one could measure
the conductance $G_R$ through the right level, which will be related to
the average population of the latter, $n_R$, by the Friedel sum rule:
$G_R=\frac{e^2}{h}\left(\frac{2|t_{R1}t_{R2}|}{|t_{R1}|^2+|t_{R2}|^2}\right)^2
\sin^2(\pi n_R)$. Thus, in the regime of continuous
switching, in addition to the two usual Coulomb-blockade peaks (of
width $\Gamma_R$) one will find a ``correlation induced
resonance'' (CIR) \cite{pl_theory,slava} of width $U T_K /|\Gamma_R-\Gamma_L|$
[Fig.~\ref{fig:friedel}(a)];
as the electrostatic coupling to the QPC increases,
$T_K$ will decrease in a Kosterlitz-Thouless fashion,
$\ln{T_K} \sim (U^*_{S}-U_{S})^{-1/2}$,
until we reach the abrupt population switching regime at
$U_{S} \ge U^*_{S}$.
From that point on the CIR disappears [Fig.~\ref{fig:friedel}(b)].
Further details will be discussed elsewhere \cite{long_version}.

The analysis presented here, while specifically tackling the
ubiquitous physics of population switching and charge sensing, is
close to earlier studies of QPTs involving two-level systems
\cite{spin_boson}, including Kondo models coupled to Ohmic baths
\cite{kamenev_gefen,impurity_qpt}.
Here we have found that PS is inherently not
abrupt, but in attempting to measure it with a third terminal (a
QPC) one may induce a QPT. The system at hand is an appealing
laboratory to modify and control at will such effects as Mahan
exciton, Anderson orthogonality catastrophe and Fermi edge
singularity \cite{mahan}. It also serves to demonstrate
the strong effect of a measuring device
on a nanoscale system.

There are several obvious extensions to our analysis.
The  absence of left-right symmetry in the original model
[Fig.~\ref{fig:geometry}(a)], implies a
finite inter-dot hopping $t_{LR}$ in the equivalent model,
Fig.~\ref{fig:geometry}(b) \cite{slava}.
This simply expresses the possible transition of an electron
from one level to the other through the leads, and will smear
the PS, as first noted in the last of Refs.~\cite{population_switching}.
Finite temperature will also have a rounding effect. Thus,
in the proposed experiment one could try to follow the manner of
decrease of the width of the CIR as a function of $U_{S}$,
before it disappears as soon as $T_K<T$.
Finally, the spinful analogue of the model can be shown to
display a similar behavior in the first Coulomb-blockade valley.
All these issues will be elaborated elsewhere \cite{long_version}.

\begin{acknowledgments}
We thank D.I. Golosov and J. von Delft for useful discussions.
Financial support from the Adams Foundation of the Israel Academy
of Science, EU project GEOMDISS, GIF, ISF (Grants 569/07, 715/08),
the Minerva Foundation, and Schwerpunkt Spintronics SPP 1285
is gratefully acknowledged.
\end{acknowledgments}

\end{document}